\documentclass[12pt,a4paper,fleqn]{article}

\usepackage{mathptmx}        
\usepackage{helvet}          
\usepackage{courier}         

\usepackage{makeidx}         
\usepackage{graphicx}        
\usepackage{multicol}        
\usepackage[bottom]{footmisc}

\usepackage{url}             

\makeindex             

\begin{document}
{ 
\def\widebreve#1{\mathop{\vbox{\ialign{##\crcr\noalign{\kern3\p@}
    \brevefill\crcr\noalign{\kern3\p@\nointerlineskip}
    $\hfil\displaystyle{#1}\hfill$\crcr}}}\limits}
\def\brevefill{$\m@th\bracelu\leaders\vrule\hfill\braceru$}
\def\widebreve#1{\mathop{\vbox{\ialign{##\crcr\noalign{\kern3\p@}
    \brevefill\crcr\noalign{\kern3\p@\nointerlineskip}
    $\hfil\displaystyle{#1}\hfill$\crcr}}}\limits}
\def\brevefill{$\m@th\bracelu\leaders\vrule\hfill\braceru$}
\title{DNS of compressible multiphase flows through the
  Eulerian approach} 
\author{Matteo Cerminara
  \\
  Scuola Normale Superiore di Pisa
  \\
  Istituto Nazionale di Vulcanologia e Geofisica
  \\
  Pisa, ITALY, matteo.cerminara@sns.it 
  \and
  Luigi C. Berselli
  \\
  Dipartimento di Matematica
  \\
  Universit\`a di Pisa
  \\
  Pisa, ITALY, berselli@dma.unipi.it 
  \and
  Tomaso Esposti Ongaro
  \\
  Istituto Nazionale di Geofisica e Vulcanologia
  \\
  Section of Pisa
  \\
  Pisa, ITALY, ongaro@pi.ingv.it
  \and
  Maria Vittoria Salvetti
  \\
  Dipartimento di Ingegneria Aerospaziale
  \\
  Universit\`{a} di Pisa
  \\
  Pisa, ITALY, mv.salvetti@ing.unipi.it
}
\maketitle
\begin{abstract}
 In this paper we present three multiphase flow models suitable for the study of the dynamics of
compressible dispersed multiphase flows. We adopt the Eulerian approach because we focus our
attention to dispersed (concentration smaller than 0.001) and small particles (the Stokes number has
to be smaller than 0.2). We apply these models to the compressible ($\textup{Ma} = 0.2,\,0.5$)
homogeneous and isotropic decaying turbulence inside a periodic three-dimensional box ($256^3$
cells) using a numerical solver based on the OpenFOAM$^\textup{\tiny\textregistered}$ \verb!C++!
libraries. In order to validate our simulations in the single-phase case we compare the energy
spectrum obtained with our code with the one computed by an eighth order scheme getting a very good
result (the relative error is very small $4*10^{-4}$).
Moving to the bi-phase case, initially we insert inside the box an homogeneous distribution of
particles leaving unchanged the initial velocity field. Because of the centrifugal force,
turbulence induce particle preferential concentration and we study the evolution of the solid-phase
density. Moreover, we do an {\em a-priori} test on the new sub-grid term of the multiphase
equations comparing them with the standard sub-grid scale term of the Navier-Stokes equations.
\end{abstract}
\section{Introduction}
This work is part of a long-term project concerning modeling, simulation and
analysis of particle-laden turbulent plumes, motivated by the study of 
the injection of ash plumes in the atmosphere during explosive volcanic eruptions. 
Ash plumes represent indeed one of the major volcanic hazards, since they can produce widespread
pyroclastic fallout in the surrounding inhabited regions, endanger aviation and convect fine
particles in the stratosphere, potentially affecting climate. 
Volcanic plumes are characterized by the ejection of a mixture of gases and polydisperse particles
(ranging in diameter from a few microns to tens of millimetres) at high velocity (100-300 m/s) and
temperature (900-1100 $^o$C), resulting in an equivalent Reynolds number exceeding $10^7$ (at
typical vent diameters of 10-100 m) \cite{V2001}. Estimates of the particle concentration in the
plume \cite{W1988} suggest that particle volume fraction decreases rapidly above the vent by the
concurrent effect of adiabatic expansion of hot gases and turbulent entrainment of air, down to
values below $10^{-3}$, at which plume density becomes lower than atmospheric density. 

To simulate such phenomenon by means of a fluid dynamic model, Direct Numerical Simulation (DNS) and
a Lagrangian description of particles are beyond our current computational capabilities.
Following \cite{TEO2007}, we therefore describe the eruptive mixture by adopting a multiphase flow
approach, i.e., solid particles are treated as continuous, interpenetrating fluids (phases)
characterized by specific rheological properties. For each phase, the Eulerian multiphase balance
equations of mass, momentum and energy are considered. A Large Eddy Simulation (LES) framework,
requiring the specification of sub-grid closure terms, is used to account for turbulence.
Even by adopting such an approach, however, the description of a large number of Eulerian
phases is still extremely computationally costly and has not allowed, so far, to perform reliable
multiphase LES of volcanic plumes.

The aim of the present work is to formulate a faster, Eulerian multiphase flow model able to
describe the most relevant non-equilibrium behaviour of volcanic mixtures (such as the effect of the
grain-size distribution on mixing and entrainment, particle clustering and
preferential concentration of particles by turbulence) while keeping the computational cost as low
as possible, in order to achieve a sufficient resolution to perform reliable LES at the full
volcanic scale.

In Section 2, we discuss a hierarchy of multiphase flow models, their potential and limits. Among
them, a quasi-equilibrium model based on a first-order asymptotic expansion of model equations in
powers of $\tau_s$ (the particle equilibrium time,~\cite{FB2001}) is preferred in the context of
volcanic plume
simulations. Section 3 focuses on the capability of such models to describe gas-particle homogeneous
and isotropic turbulence in compressible regime.
Finally, we discuss the properties of the Favre-filtered models and the a-priori estimates
of sub-grid terms, which is preliminary to the formulation of a closure model for LES of
volcanic plumes.

\section{Eulerian multiphase flow models}
In order to use Eulerian models, we first discuss the physical constraints that characterized
volcanic plumes.
1) Grain size distribution can be discretized into a finite number of particulate classes.
2) Particles are heavy, i.e., 
${\hat{\rho}_s}/{\hat{\rho}_g} \gg 1$ ($\hat{\rho}_s \approx 400-3000$
kg/m$^3$), where the hat denotes the density of the phase material.
3) Each class can be described as a continuum (Eulerian approach), i.e., the mean free path is much smaller than, say, the numerical grid size). 
4) Low concentration: particle volume fraction
$\epsilon_s=V_s/V <10^{-3}$. Under such conditions, particles can be
considered as non-interacting (pressure-less) or weakly-interacting (small pressure term). 
 As a consequence of 2) and 4), the bulk densities
($\rho_g=(1-\epsilon_s)\hat{\rho}_g$ and
$\rho_s=\epsilon_s\hat{\rho}_s$) can be of the same order of
magnitude (about 1 kg/m$^3$ or less).

Following the same approach described in ~\cite{BE2010}, we can build a hierarchy of models based on
the ratio between a particle characteristic equilibrium time $\tau_s$ and a characteristic time of
the fluid flow $\tau_{\eta}$ (which can be the Kolmogorov time in the case of developed turbulent
flows, or a large eddy turnover time). 

 Moreover, to simplify our analysis, we will assume in the following that the flow is
iso-entropic (i.e., we neglect the initial explosive phase, which affects only the first hundreds of
metres above the vent), and we consider a barotropic model, thus avoiding to solve the full energy
equation. 
 This approach is widely used in atmospheric and plume models and verified by experiments
\cite{M1956,W1988}.\\

\noindent\textbf{Barotropic Eulerian-Eulerian model: }
For small relative Reynolds number,\\
 ${\rho_g|\mathbf{u}_s-\mathbf{u}_g|d_s}/{\mu}<1$ (where $d_s$ is the
 particle's diameter, $\mu$ and $\rho$ are gas density and dynamic viscosity and $\mathbf{u}_g$ and
$\mathbf{u}_s$ are the gas and solid phase velocities) the drag force between
the gas and a solid phase can be expressed by the Stokes law
 $\mathbf{f}_d={\rho_s}(\mathbf{u}_s- \mathbf{u}_g)/{\tau_s}$ and particle relaxation time is
defined by $\tau_s={\hat{\rho}_s d_s^2}/{18 \mu}$. 
 
When $\tau_s$ is of the same order of magnitude as $\tau_{\eta}$, the fully coupled multiphase flow
equations must be considered. For a two-phase (monodisperse) mixture, the system of mass and
momentum
balance equations reads:
     \begin{eqnarray}
&       \partial_t\rho_g + \nabla\cdot(\rho_g \mathbf{u}_{g}) = 0,
\qquad       \partial_t\rho_s + \nabla\cdot(\rho_s \mathbf{u}_{s}) =0, 
       \\
 &      \partial_t (\rho_{g} \mathbf{u}_{g}) 
      +\nabla\cdot (\rho_g
       \mathbf{u}_{g}\otimes \mathbf{u}_{g} + p_g\,\mathrm{I}) -\nabla\cdot\sigma 
       = \frac{\rho_{s}}{\tau_{s}}(\mathbf{u}_{s}-\mathbf{u}_{g}) + \rho_g \mathbf{g},
       \\
&       \partial_t (\rho_{s} \mathbf{u}_{s}) +\nabla\cdot
       (\rho_s \mathbf{u_{s}}\otimes \mathbf{u_{s}} + p_s\,\mathrm{I})   -
       \nabla\cdot \pi
       = -
       \frac{\rho_{s}}{\tau_{s}}(\mathbf{u}_{s}-\mathbf{u}_{g})     + \rho_s \mathbf{g},
\end{eqnarray}
where the subscripts $g$ and $s$ represent the gas and solid phase, respectively,
 the pressure terms $p_s$ and $p_g$ are given by the barotropic model as 
$p=p_{0}({\rho}/{\rho_{0}})^{\gamma}$
(but the exponent $\gamma$ may be different),
$\pi=\mu_{s}\rho_{s}\,\nabla^{sym}\mathbf{u}_{s}$, 
$\sigma=\mu\left(\nabla^{Sym}\mathbf{u}_{g} -\frac{2}{3}
    \nabla\cdot\mathbf{u}_{g}\,\mathrm{I}\right) $ and $\gamma$ ($\gamma_s$) is the
  specific heat ratio of the gaseous (solid) phase. Here $\mathbf{g}$ is the gravitational
acceleration, which has not been considered in the following simulations.\\

\noindent\textbf{Fast Eulerian model:} for smaller (but non negligible)
$\tau_s$, it is possible to insert a first-order approximation of
the particle velocity into the Eulerian-Eulerian equations:
$\mathbf{u}_s=\mathbf{u}_g-\tau_s\mathbf{a}_{g}$, where
$\mathbf{a}_{g}=\frac{D}{Dt}(\mathbf{u}_{g})-\mathbf{g}$ is the gas phase acceleration
(cf.~\cite{FB2001}).

This allows to reduce the number of model equations to one single momentum equation for the
mixture plus N equations for mass conservation.
By assuming also local thermal equilibrium, we get the following system, again written for a
two-phase mixture for the sake of simplicity (here $\rho_m \equiv \rho_g + \rho_s$):
\begin{eqnarray}
  \label{eq:1}
  &         \partial_t\rho_g + \nabla\cdot(\rho_g \mathbf{u}_{g}) = 0,
  \qquad   \partial_t\rho_s + \nabla\cdot(\rho_s \mathbf{u}_{s}) =0,
  \\ 
  &   \partial_t (\rho_m \mathbf{u}_g) + \nabla\cdot (\rho_m \mathbf{u}_g\otimes
  \mathbf{u}_g + p\mathrm{I}) - \nabla\cdot\sigma = \rho_m
  \mathbf{g}+\tau_{s}\nabla\cdot(\rho_s \mathbf{a}_{g})\mathbf{u}_{g}. 
  \end{eqnarray}
%
%
%
%

\noindent\textbf{Dusty Gas model:} 
Finally, when $\tau_s$ is negligible, the fast Eulerian model reduces to the so-called dusty-gas
model \cite{Mar1970}, in which all phases have the same velocity.
As predicted by the theory, preliminary simulations carried out for turbulent plumes have shown that
the dusty-gas model is not capable of reproducing preferential concentration and the effect of
particle inertia on turbulent mixing. Therefore, such a model  does not appear to be well suited to
our purposes, and, hence, the results of the dusty-gas model will not be shown herein.

We implemented these models into OpenFOAM$^\textup{\tiny\textregistered}$.
Since the flow density is varying in time and space due to the presence of particles, we adopted
a robust numerical scheme to simultaneously treat compressibility, buoyancy effects, and turbulent
dispersal dynamics, which is based on a segregated solution algorithm~\cite{D1993}, with an adaptive
time-stepping. Since we limit our analysis to moderate $\textup{Ma}_\textup{rms}$ (up to 0.6), we do
not need shock-capturing schemes. Here $(\cdot)_\textup{rms}$ stands for root mean square.

\section{Numerical results}
As a first step for the appraisal of the considered models, we performed numerical tests
by performing DNS of decaying homogeneous isotropic turbulence as reported
in~\cite{GMSCD1999,LOC2005}.
We consider an initial solenoidal velocity field with spectrum
$E(k)\propto({k}/{k_0})^4\exp\big(-2\,({k}/{k_0})^2\big)$, $k_0=2$ and viscosity $\mu \simeq
0.00239$ to have the Taylor microscale $\lambda=1/2$, $Re_\lambda=50$ and that the
Kolmogorov scale $\eta$ remains bigger than the mesh-size: $\eta/\Delta x \geq 2.0$. The
initial velocity field is identical for both the gas and the solid, the
Stokes number at $t=0$ is $St=0.02$, the specific-heat ratios are
$\gamma=1.4$ and $\gamma_s=1$ and the solid phase viscosity $\mu_s=10^{-8}$.
Since $\rho_s=\rho_g=1$, the inertial forces of particles are significant in the flow dynamics.

To validate our simulations in the single-phase case, we compared the energy
spectrum obtained with our code with the one computed by an eighth order scheme
\cite{PG2004}. The comparison is made after one large-eddy turnover time at
$\textup{Ma}_\textup{rms}=0.2$, we find the $L^2$ norm of the difference between the two spectra is
$4.0*10^{-4}$ (cf. Fig.~\ref{fig:energySpectrum}). This validates the accuracy of our numerical
solver in the single-phase case.

As in~\cite{GMSCD1999}, we keep the same velocity field in all simulations and modified the initial homogeneous pressure field in order to modify the Mach number. In the
single-phase case we used $\textup{Ma}_\textup{rms}=0.2,\,0.5$. In Fig.~\ref{fig:energySpectrum} we
compare the two resulting energy spectra. In the gas-particle case, we keep the same velocity and
pressure fields, modifying in this way the non dimensional
parameters. In particular the equivalent Reynolds number doubles because the mixture bulk density
moves from $\rho_g = 1$ to $\rho_g+\rho_s =2$, while the equivalent Mach number increases of a
factor $\simeq 1.4$ because the speed of sound of the mixture decreases.

\begin{figure}[h]
  \centering
       \includegraphics[width=0.45\columnwidth]{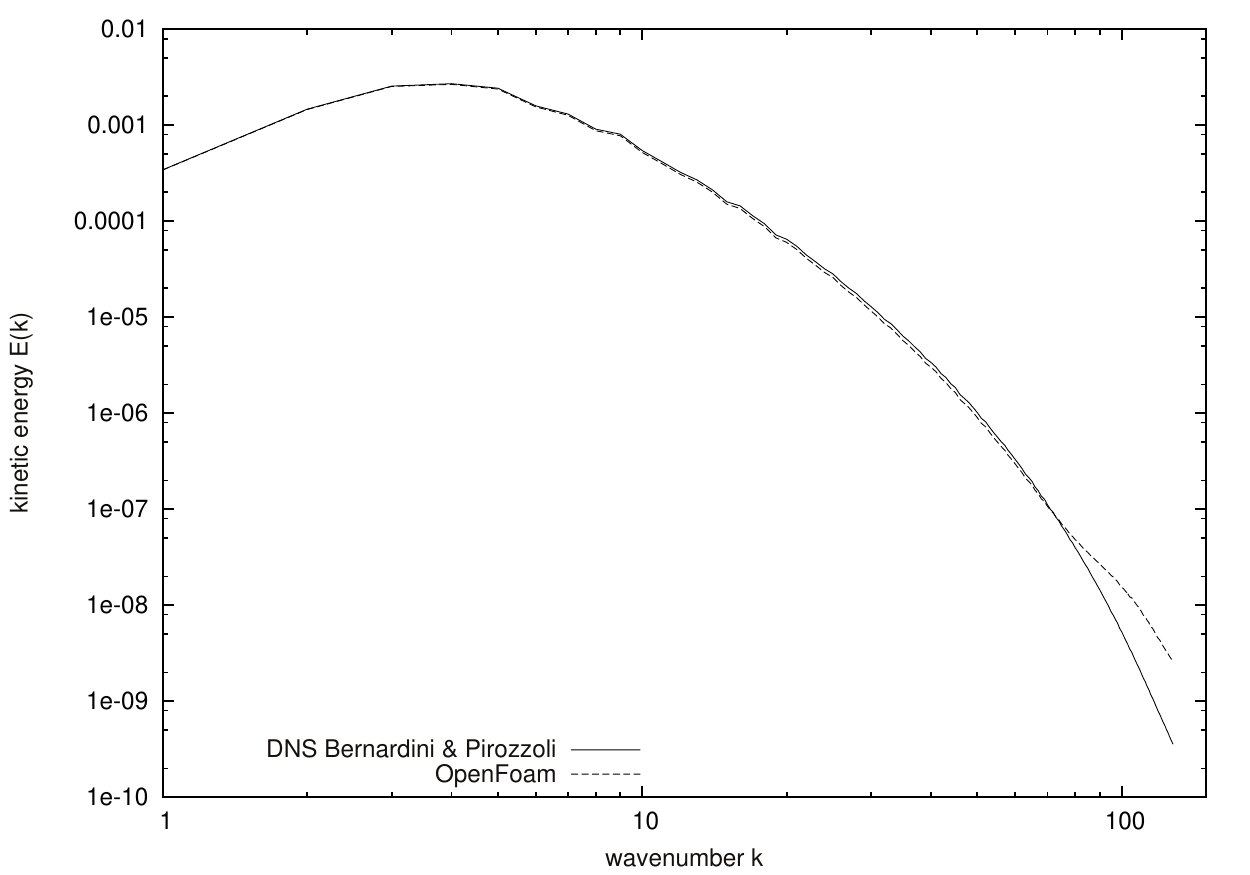}
     \includegraphics[width=0.45\columnwidth]{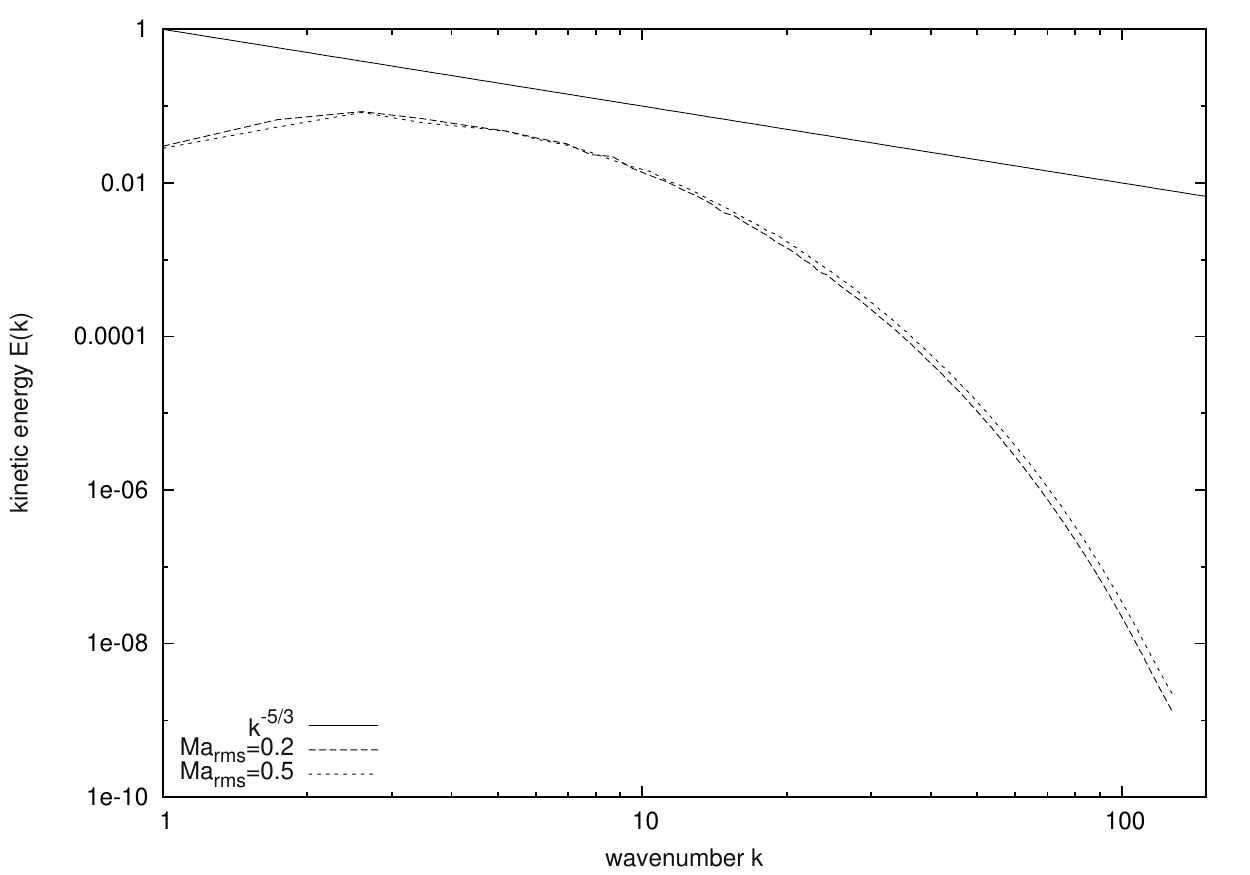}

     \caption{Comparison of the Energy spectrum: PISO \textit{vs.}
       eight order DNS (see~\cite{PG2004}) after one large-eddy
       turnover time (left).  PISO at
       $\textup{Ma}_\textup{rms}=0.2,\,0.5$ after two large-eddy
       turnover time (right).}\label{fig:energySpectrum}
\end{figure}

\noindent\textbf{Comparison of the two Eulerian models:}
Fig~\ref{fig:globalTurb} shows the time evolution of kinetic energy and
enstrophy obtained in the single-phase case (NSE) and in the two-phase one by the barotropic
Eulerian and fast Eulerian models respectively.
\begin{figure}[h]
  \centering
       \includegraphics[width=0.45\columnwidth]{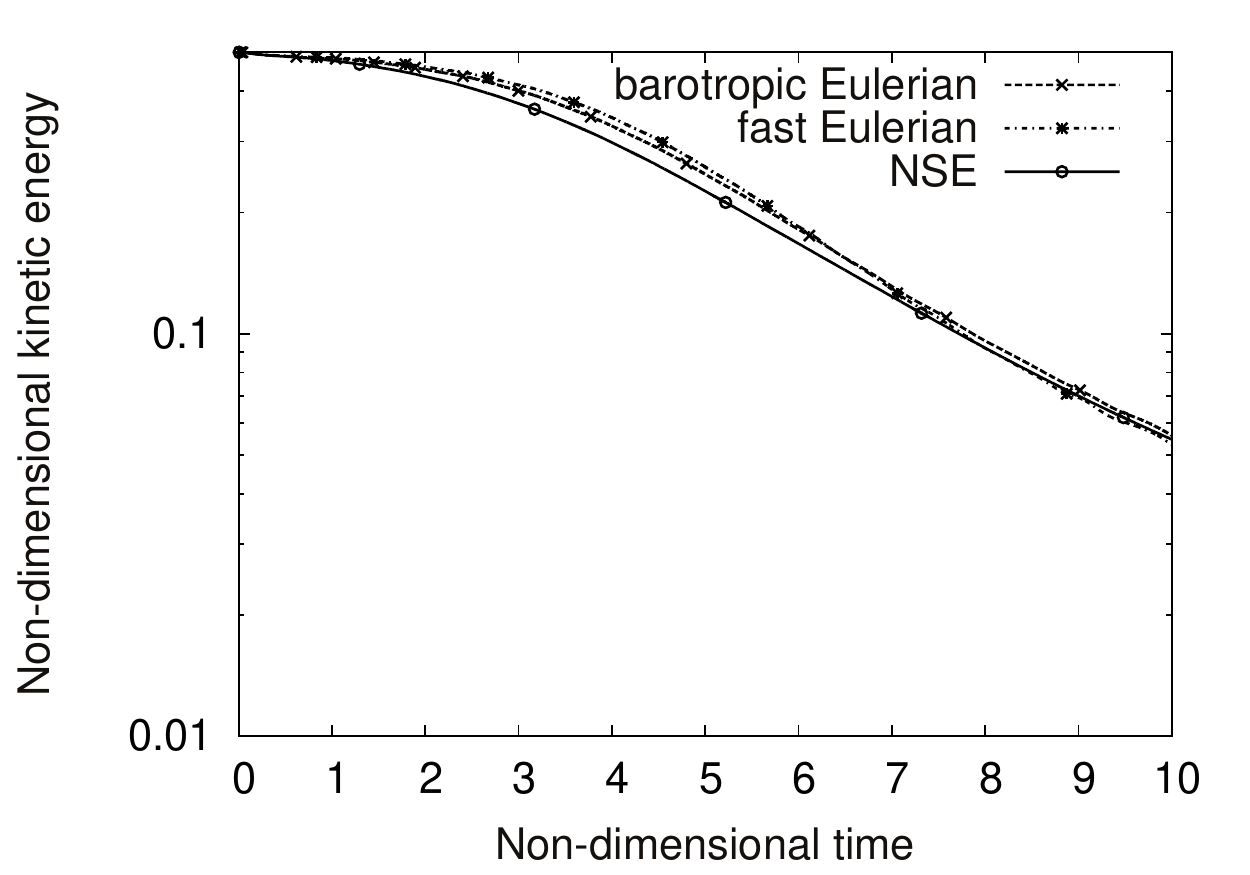}
     \includegraphics[width=0.45\columnwidth]{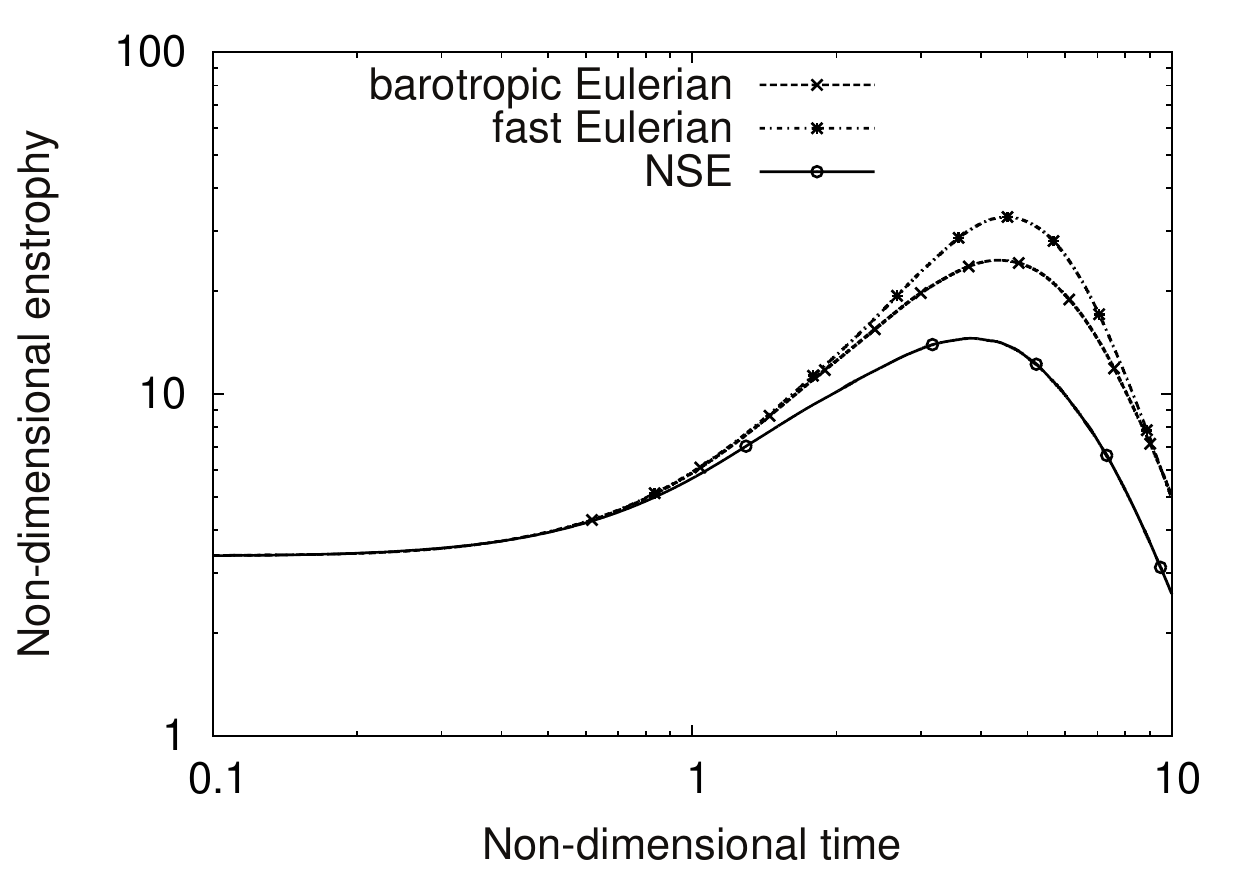}
 \caption{Turbulence global properties evolution: Kinetic energy
   (left) and Enstrophy (right) .
}\label{fig:globalTurb}
\end{figure}
We first observe that, as expected because of the doubled
$\textup{Re}$, the vorticity and the energy are larger than in
the single-phase case. Next, we observe that in our codes the Fast Eulerian model
is less diffusive than the Eulerian one (the enstrophy becomes larger and the
turbulent dissipation is more efficient, cf.~\cite{GMSCD1999}). This fact is qualitatively
observable also in Fig.~\ref{fig:snapshot}, where we show isosurfaces of the solid density for
the two Eulerian models.
\begin{figure}[h]
  \centering
       \includegraphics[width=0.95\columnwidth]{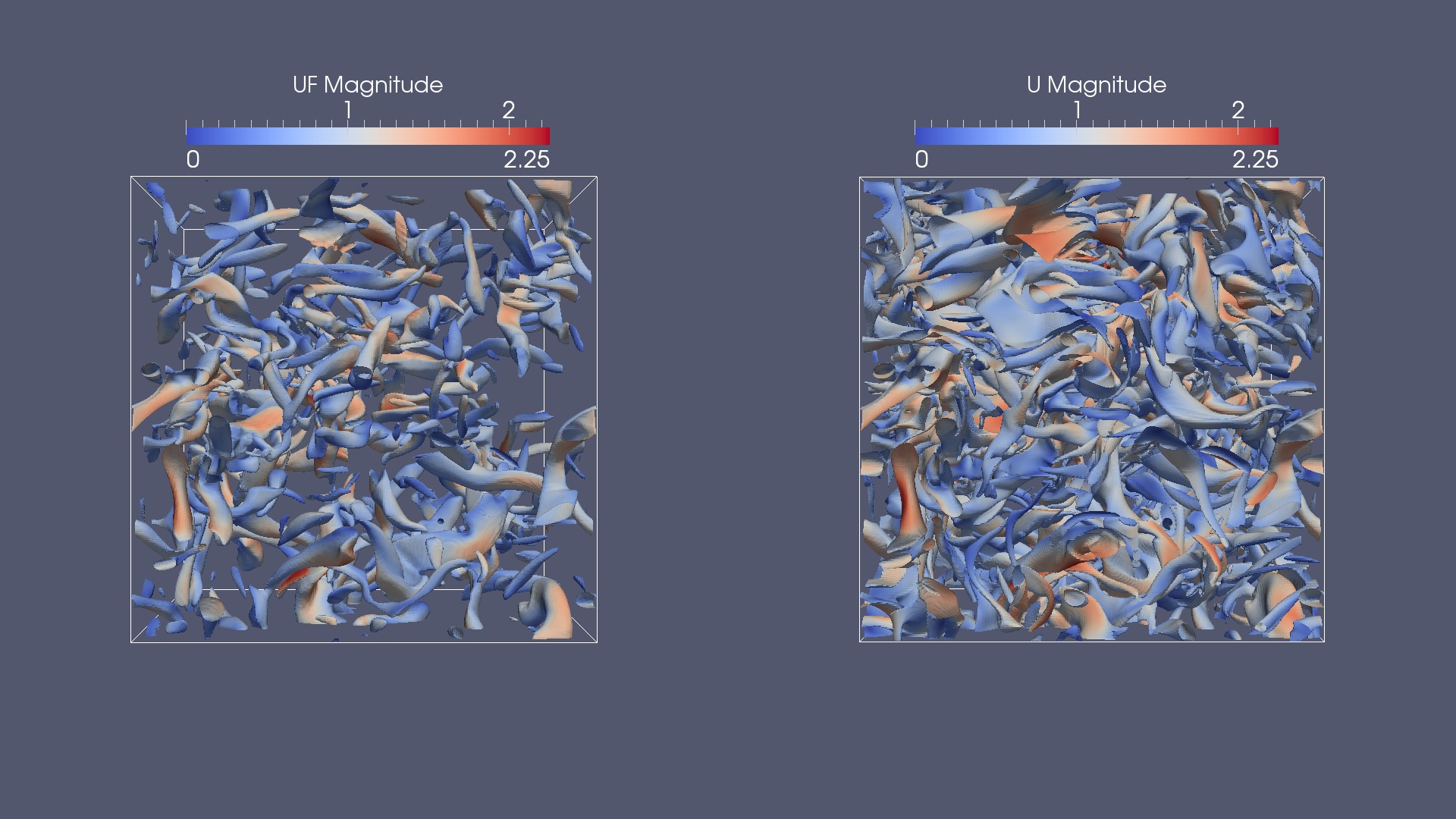}
 \caption{Isurfaces of the solid phase density: barotropic Eulerian
   (left) and fast Eulerian (right). Turbulence induces particle preferential concentration.
}\label{fig:snapshot}
\end{figure}
In Fig.~\ref{fig:densities} we report the evolution of the density
fluctuations, for both the solid and gaseous phases.
\begin{figure}[h]
  \centering
       \includegraphics[width=0.45\columnwidth]{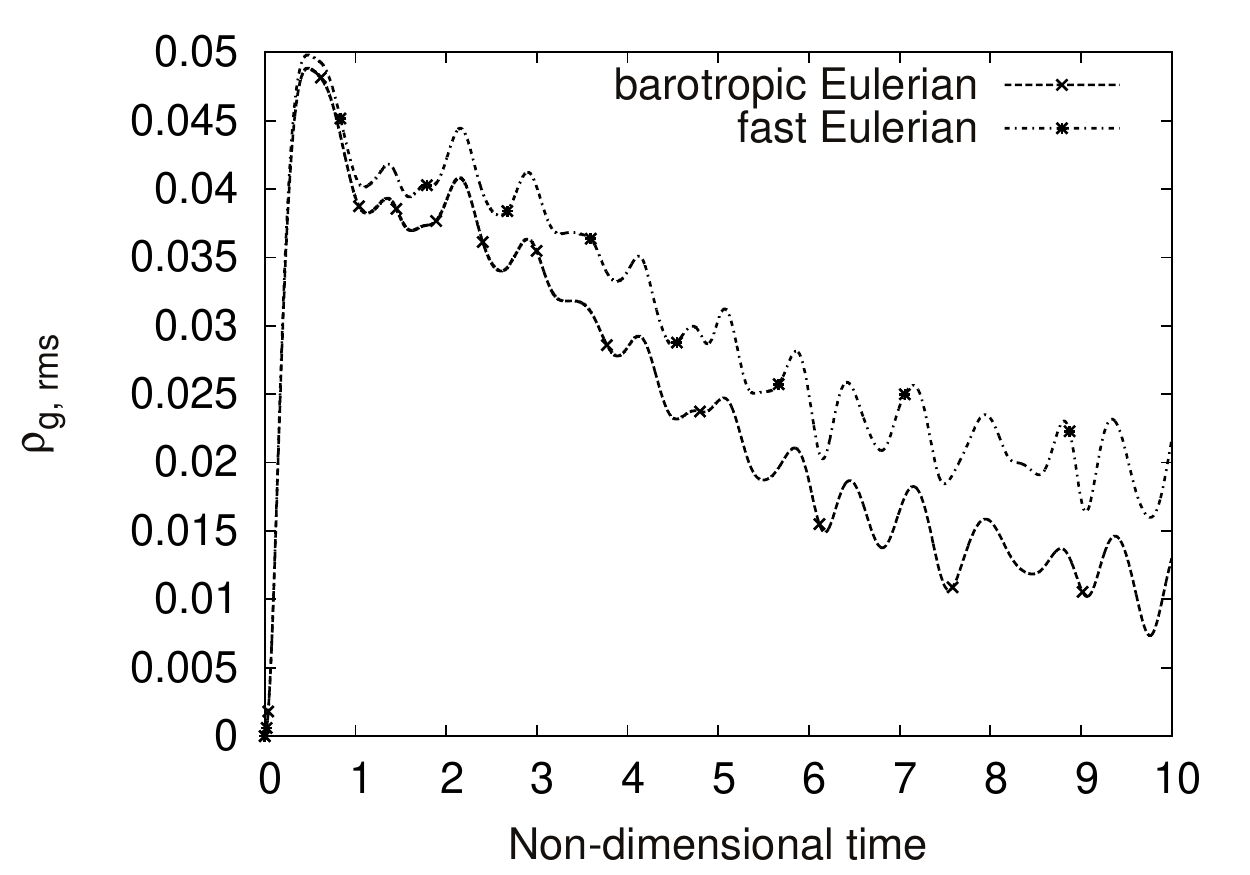}
       \includegraphics[width=0.45\columnwidth]{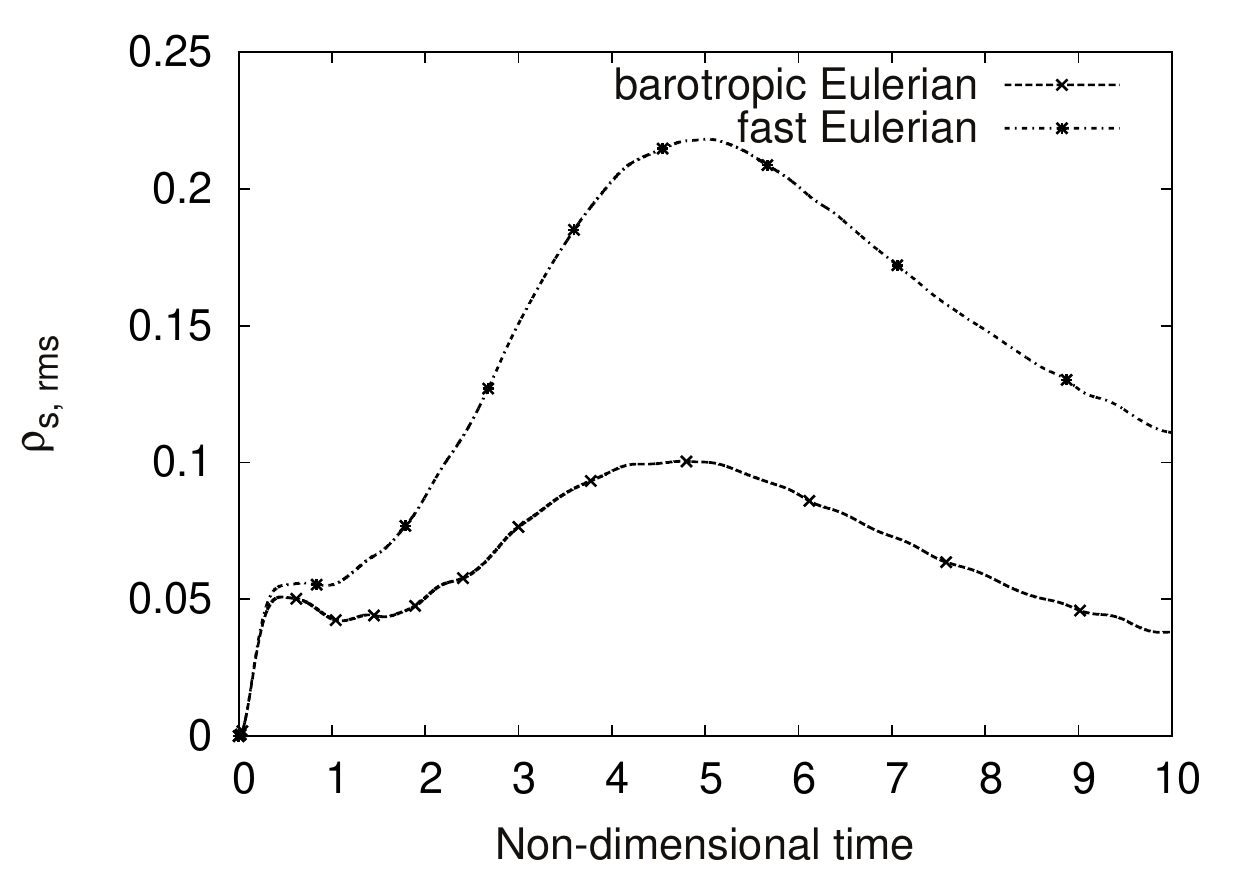}
       \caption{Density fluctuations $\sqrt{\langle(\rho -
           \langle\rho\rangle)^2\rangle}$: Gaseous phase $\rho_g$
         (left), solid phase $\rho_s$ (right).  }\label{fig:densities}
\end{figure}
We find that the presence of a massive solid phase increases
the density fluctuation (the Mach number is modified by the
presence of a solid phase), and that in the Fast Eulerian model the preferential
concentration of the solid particles is stronger than in the Eulerian
simulation (the latter being more diffusive than the former). These differences between the two
models are quite surprising since for the values of the Stokes number of the considered particles
they are expected to give close results. A possible cause may be a difference in
the treatment of source term in the two Eulerian models.
In particular, the Stokes coupling between the two phases has been treated explicitly in the
Eulerian model, probably underestimating its fluctuations.
This will be further investigated in future works.
\\

\noindent\textbf{A priori tests:} Moving forward, we use the DNS results to evaluate the SGS terms
which would arise in LES.
To filter model equations, we have defined
a Favre filter $\tilde{\phantom{a}},\,\breve{\phantom{a}}$, for
each phase so that: $ \overline{\rho_g \psi} = \overline{\rho_g}
\tilde{\psi}$ and $\overline{\rho_s \psi} = \overline{\rho_s}
\breve{\psi}$.
    Using these filters in our models
    we get new sub-grid terms, which are evaluated on the DNS results.

    In the barotropic Eulerian model, the
    SGS terms different from zero are:
$$
\textup{E}_1= \nabla\cdot \tau, \quad \textup{E}_2=\nabla\cdot
\theta,\quad
\textup{E}_3={\overline{\rho_{s}}}(\breve{\mathbf{u}}_{g}-\tilde{\mathbf{u}}_{g})/{\tau_s},\quad
\textup{E}_4=a \nabla
(\overline{\rho_{g}^\gamma}-\overline{\rho_{g}}^\gamma),
$$
where $a=p_0/\rho_0^\gamma$, $ \tau= \bar{\rho}_{g}
(\widetilde{\mathbf{u}_{g}\otimes \mathbf{u}_{g}} -
\widetilde{\mathbf{u}}_{g}\otimes \widetilde{\mathbf{u}}_{g})$, and
$\theta = \bar{\rho_{s}} (\widetilde{\mathbf{u}_{s}\otimes
  \mathbf{u}_{s}}-\breve{\mathbf{u}}_{s}\otimes
\breve{\mathbf{u}}_{s})$. The subgrid terms $\textup{E}_1$ and $\textup{E}_2$ represent the
divergence of the  classical SGS stress tensor for the gas and the solid phases respectively, while
$\textup{E}_3$ represents the SGS effects on the Stokes drag acting on the particles and
$\textup{E}_4$ the barotropic pressure SGS term. Figure 2 shows the time evolution of the r.m.s. of
$\textup{E}_i$ in semi-log scale. The filtering is made by a
top hat filter with radius $\delta=\sqrt{2}h$.

\begin{figure}
\centering
\includegraphics[width=0.49\columnwidth,keepaspectratio=true]{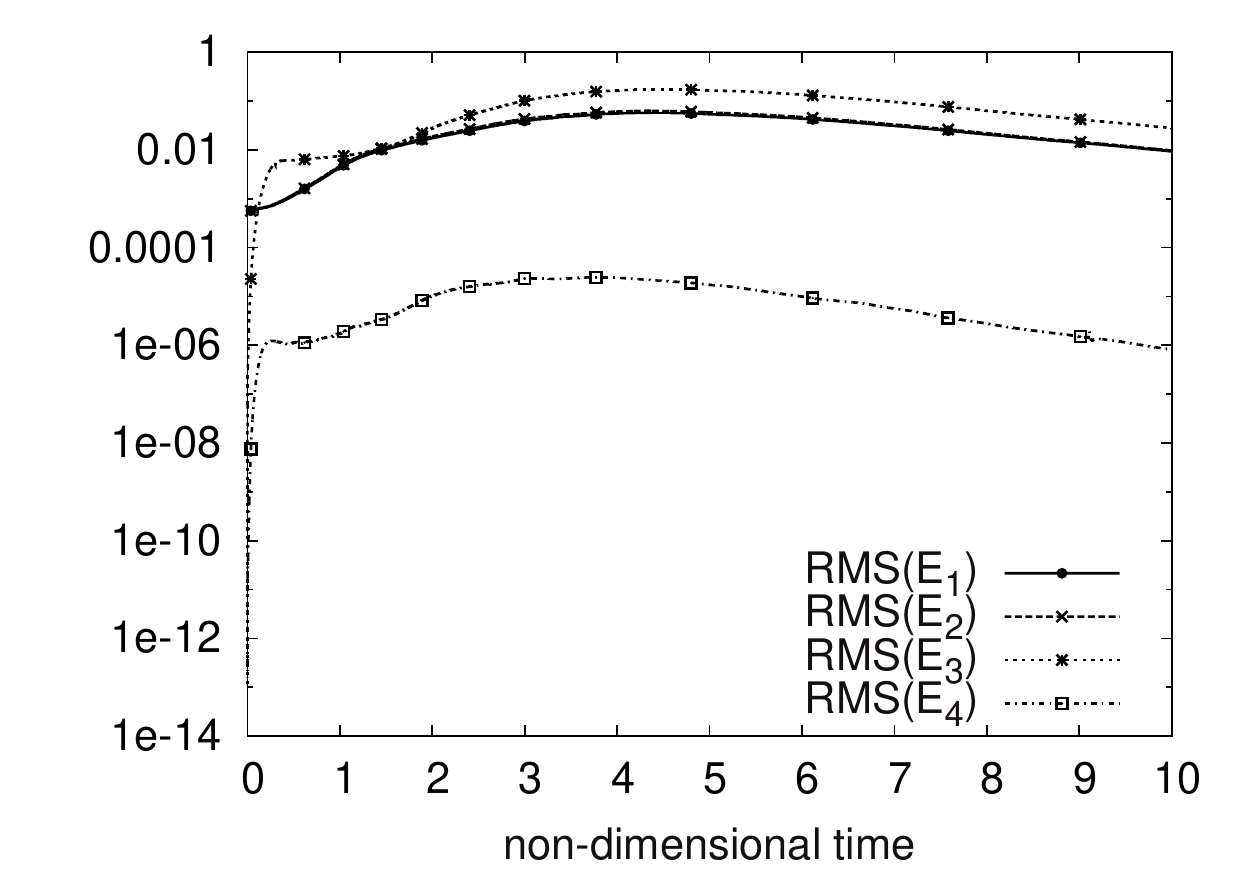}\,
\includegraphics[width=0.49\columnwidth,keepaspectratio=true]{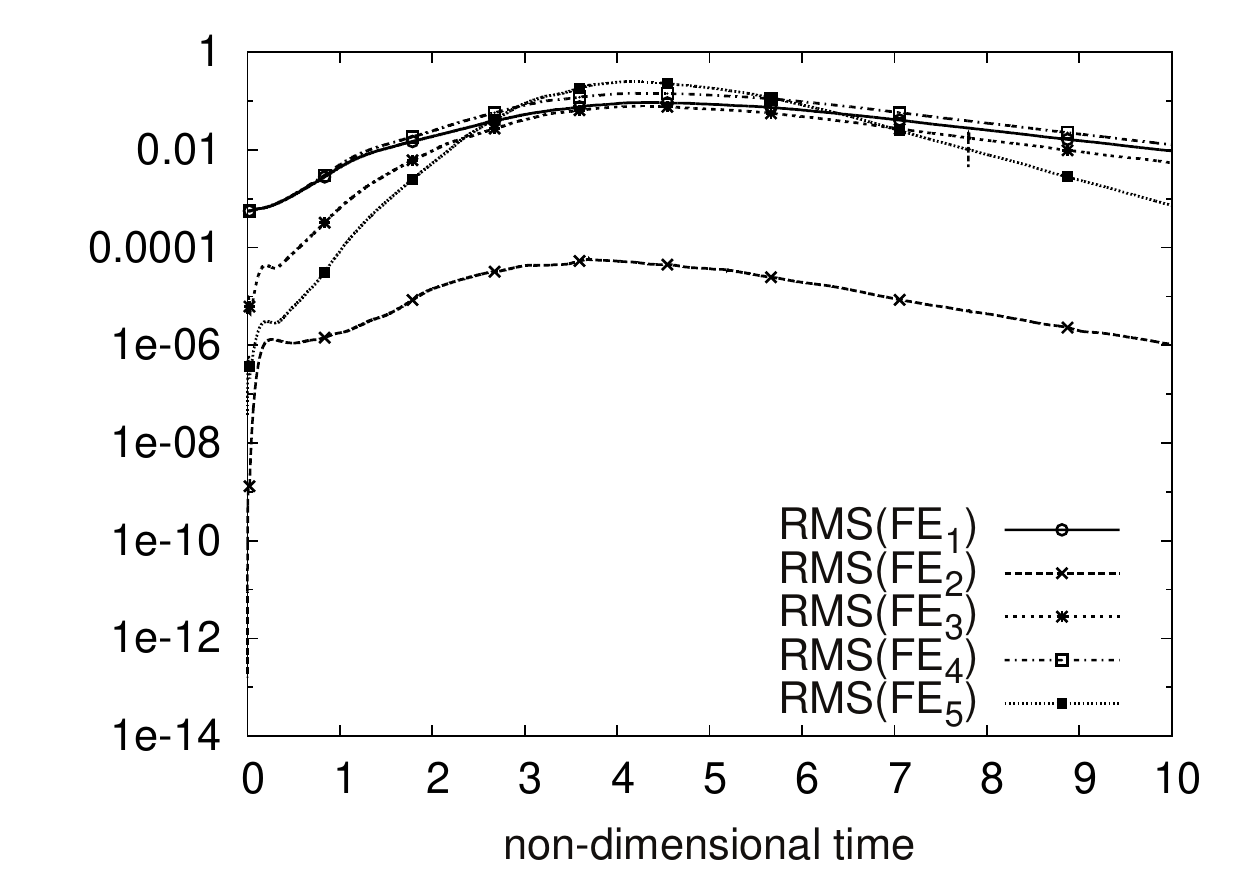}
\caption{ Barotropic Eulerian: Evolution of the  RMS of $\textup{E}_i$ at
  Mach=0.35 (left), 
 Barotropic Fast Eulerian: Evolution of the  RMS of $\textup{FE}_i$ at Mach=0.28 (right).}
\end{figure}
We observe that the subgrid term related with the barotropic pressure
is significantly smaller than the others. Hence, in addition to the
terms present in the mono-phase case, also the Stokes force of
interaction between the two phases is of the same order of magnitude
and requires LES modeling.

In the case of the Fast Eulerian model, different terms comes out from the Favre
filtering operation. In particular, we have:
\begin{eqnarray*}
  &\textup{FE}_1=\nabla\cdot \bar{\rho}_{g}(\widetilde{\mathbf{u}_{g}    \otimes
    \mathbf{u}_{g}} -    \tilde{\mathbf{u}}_{g}    \otimes
  \tilde{\mathbf{u}}_{g}),
  \ \ 
  \textup{FE}_2=\overline{p}-p_{0}(\overline{\rho}_{g}/{\rho_{0}})^{\gamma},
  \\ 
  &\textup{FE}_3=\partial_{t}(\overline{\rho}_{s}(\breve{\mathbf{u}}_{g}-\tilde{\mathbf{u}}_{g})),
  \ \ 
  \textup{FE}_4=\nabla\cdot \bar{\rho}_{g}(\breve{\mathbf{u}_{g}    \otimes
    \mathbf{u}_{g}} -    \tilde{\mathbf{u}}_{g}    \otimes
  \tilde{\mathbf{u}}_{g}),
  \\ 
  &\textup{FE}_5=\tau_{S}\left[\overline{\nabla\cdot(\rho_{s}\mathbf{a}_{g})\mathbf{u}_{g}}-
    \nabla\cdot(\overline{\rho}_{s}\tilde{\mathbf{a}}_{g})\tilde{\mathbf{u}}_{g}\right]   
\end{eqnarray*} 
As in the other case the subgrid term related with the barotropic
pressure is significantly smaller than the others. 
The other terms require LES modeling and we can observe that they are
of the same order of magnitude of the relevant terms appearing in the
Eulerian case.
\section{Conclusions}
We performed DNS of multiphase flows comparing results between the barotropic Eulerian and
the fast Eulerian model. Clearly the fast Eulerian one is less time consuming (about a factor one
half), but it seems that at least in our setting the overall quality of the
numerical results is still acceptable. The fast Eulerian model seems to be less
diffusive and consequently the preliminary results concerning the
preferential concentration are slightly better. A conclusive
assessment requires comparison with Lagrangian simulations (as those
in~\cite{BCDLM2007}), which we do not have currently at disposal, due
to the peculiarity of the large initial solid phase inertia. Further
tests are ongoing with the objective of well reproducing with LES both
turbulence and preferential concentration for poly-disperse mixtures.

}

\end{document}